\documentclass[preprint2]{aastex}

\begin{document}

\title{Structure and Star Formation in NGC 925\footnote{to appear in the August
2000 Astronomical Journal}}
\author{D.J. Pisano}
\affil{Dept. of Astronomy, University of Wisconsin}
\affil{475 N. Charter St., Madison, WI 53706}
\email{pisano@astro.wisc.edu}
\author{Eric M. Wilcots}
\affil{Dept. of Astronomy, University of Wisconsin}
\affil{475 N. Charter St., Madison, WI 53706}
\email{ewilcots@astro.wisc.edu}
\author{Bruce G. Elmegreen}
\affil{IBM Research Division, T.J. Watson Research Center}
\affil{P.O. Box 218, Yorktown Heights, NY 10598}
\email{bge@watson.ibm.com}

\begin{abstract}
	We present the results from an optical study of the stellar \& star formation
properties of NGC 925 using the WIYN 3.5m telescope.  Images in B,V,R, \& H$\alpha$ 
reveal a galaxy that is fraught with asymmetries.  From isophote fits we discover that
the bar center is not coincident with the center of the outer isophotes nor with the
dynamical center (from Pisano {\it{et al.}} 1998).  Cuts across the spiral arms reveal
that the northern arms are distinctly different from the southern arm.  The southern
arm not only appears more coherent, but the peaks in stellar and H$\alpha$ emission are 
found to be coincident with those of the H~I distribution, while no such consistency is 
present in the northern disk.  We also examine the gas surface density criterion for 
massive star formation in NGC 925, and find that its behavior is more consistent with that 
for irregular galaxies, than with late-type spirals.  In particular, star formation persists
beyond the radius at which the gas surface density falls below the predicted critical value
for star formation for late-type spirals.  Such properties are characteristic of 
Magellanic spirals, but are present at a less dramatic level in NGC 925, a late-type spiral. 
\end{abstract}


\section{Introduction}

        This paper is part of an ongoing series of papers studying the general
H~I and optical properties of late-type barred spiral galaxies (Sbc-Sd, hereafter LTBS).  
We will examine the stellar and star forming properties of NGC 925 based on observations
from the WIYN 3.5m telescope{\footnote{The WIYN observatory is a joint facility of 
the University of Wisconsin-Madison, Indiana University, Yale University, and 
the National Optical Astronomy Observatories.}}.  We previously observed NGC 925
and NGC 1744, both LTBS, in H~I as part of a study of the gaseous properties of
this class of galaxies (Pisano {\it{et al.}}, 1998, hereafter Paper I), and to
determine the pattern speed of their bars (Elmegreen {\it{et al.}} 1998).  NGC 925
was chosen for study because it is a nearby ($\sim$9.3 Mpc, Silbermann {\it{et al.}} 1996),
apparently prototypical LTBS which is well oriented in inclination for dynamical and
morphological study.  This value for the distance will be used throughout this paper.  

In paper I, we found that NGC 925 had a weak spiral pattern and bar as indicated by
the small streaming motions in the spiral arms and bar.  Asymmetries are quite prevalent
in NGC 925.  NGC 925's southern spiral arm has reasonably coherent streaming motions, 
while the northern arms do not.  The center of the bar is also slightly offset from
the dynamical center of the galaxy by $\sim$950 pc.  These asymmetries could 
possibly be related to the presence of a small H~I cloud (M$_{H~I}\sim$10$^7$M$_{\odot}$)
interacting with the main galaxy.  The strength of the interaction, however, is most likely not
enough to drive the asymmetries in NGC 925.  The presence of an off-center bar, and a
single coherent spiral arm are typically viewed as characteristic properties of barred
Magellanic (SBm) galaxies (de Vaucouleurs \& Freeman 1972), and have been shown to be potentially 
long-lasting (Levine \& Sparke 1998, Noordermeer {\it{et al.}} 2000)  While NGC 925 does not illustrate
these properties as dramatically as the LMC, for example, it certainly indicates that these
characteristic asymmetries do not suddenly appear in SBm's, but are rather a
continuous change in morphology throughout late-type disk galaxies.  
A summary of basic properties of NGC 925 is given in table 1.

LTBS are an understudied and, hence, a poorly understood class of galaxies.  While
bars may play an important role in galaxy evolution by generating gas inflow 
(e.g., Matsuda \& Nelson 1977, Athanassoula 1992), driving spiral structure 
(e.g., Sanders \& Huntley 1976; James \& Sellwood 1978), and flattening the radial 
abundance gradient (Martin \& Roy 1994), these effects are shown to exist 
primarily in early-type barred spirals (S0-Sb).  LTBS do not appear to have 
strong enough bars to cause such effects (see paper I).  Previous studies of 
LTBS have found that late-type bars tend to have exponential light profiles, as opposed
to flat profiles, and end well within corotation, as opposed to just inside it (Elmegreen \&
Elmegreen 1985, Elmegreen {\it{et al.}} 1998).  Furthermore, late-type bars are not likely
to be driving the spiral pattern (Sellwood \& Sparke 1988), in contrast to early-type
bars (Elmegreen \& Elmegreen 1989).  Star formation properties are also different, with
late-type bars tending to be gas-rich and have star formation throughout the bar, while 
early-type bars have star formation only near the ends of the bar (Phillips 1995).  Our 
observations in paper I of NGC 925 provided no evidence to contradict any of these properties.

In this paper, we will investigate the optical properties of NGC 925, with a specific eye
towards the nature of star formation in the galaxy.  We will see if there are other signatures of 
asymmetry in NGC 925 that are typically associated with SBm's.  Finally, we will try to devine
what role the bar plays in determining the properties of the galaxy as a whole.  
In section 2 of this paper we discuss the observations and reduction.  In section 3, we discuss the 
optical properties of galaxy as a whole.  Section 4 discusses the star formation properties of 
NGC 925 both across the entire galaxy, and in the bar and spiral arms specifically.  
We conclude in section 5.

\section{Observations and Reductions}

        We observed NGC 925 with the WIYN 3.5m telescope during the nights of October 
6 \& 7, 1996.  The conditions were superb;  all three nights had subarcsecond seeing and 
were photometric.  We took the images through the Harris B, V, and R broadband filters 
and on and off band H$\alpha$ filters using the WIYN imager, a thinned 
2048$^2$ STIS CCD with 21$\mu$m pixels.  The pixels correspond to 
$0.196^{\arcsec}$ on the sky.  The gain was 2.8 e$^-$ per ADU, and the 
read noise was 8 e$^-$.  The imager has a total field of view of 
6.5$^{\arcmin}$.  Since NGC 925 has a major axis of approximately 
10$^{\arcmin}$, we had to mosaic two images to get full coverage of the 
galaxy.  We observed the east and west sides of the galaxy separately with 
overlap in the bar region.  The parameters of the observations are listed in 
table 2.

        We reduced the data in the usual manner.  We took bias images 
during the afternoon before each night.  We visually inspected the biases, and 
statistically compared them and found them to be identical for all three 
nights.  This being the case, we combined all 44 bias frames into one 
image for bias subtraction to maximize the signal to noise.  Each afternoon, 
we took $\sim$5 well exposed (10000-35000 counts) dome-flats per filter.
We discarded any flats which were significantly deviant from the average.  
We averaged the remaining flats together to correct for the response of the CCD.  
We switched filter wheels during each night's observations which 
sometimes caused the dust features to change in location and/or intensity.  As some of the flats 
changed between nights, we only used the flats from the same night to flatten the program 
images.  Aside from the off-line H$\alpha$ images, all images were flat to better than 5\%.  The
off-line H$\alpha$ images were flat to 10\%.  The poor quality of the flat-fielding for the
H$\alpha$-off filter may require sky-flats to improve.  

        To flux calibrate our images we observed broadband standards in the PG0231+051
and SA92 fields from Landolt (1992), and G191B2B, a spectrophotometric standard 
from Massey {\it{et al.}} (1988).  The photometric solutions for the 
broadband observations were found using the {\it{apphot}} package in IRAF.  
They are as follows:
\begin{equation}
V = OB_V + 1.08 + 0.17\times X - 0.04\times (B-V)
\end{equation}
\begin{equation}
B = (OB_{(B-V)}+OB_V) + 1.28 + 0.22\times X -0.04 \times (B-V)
\end{equation}
\begin{equation}
R = (OB_V - OB_{(V-R)}) + 0.95 + 0.10\times X -0.04\times (V-R)
\end{equation} 
where OB is the instrumental magnitude and X is the airmass.
We then applied these solutions to the bias subtracted, flat-fielded, sky 
subtracted 
images using IRAF to produce surface brightness maps.  The errors on these 
equations are roughly 0.05 mag.  As this is comparable to our flat-fielding 
errors, we combine the errors to obtain our total error on the broadband magnitudes, 
which is approximately 0.07 mag.  

        Sky subtraction was difficult given the small amount of sky present in 
our images.  There was typically only a small corner of the CCD frame which 
appeared to contain mostly sky emission.  For each side of the galaxy we used {\it{imstat}} to determine 
a mean sky level, and subtracted it from the image before mosaicing.  This value 
is quite uncertain given the lack of sky present in the images.  Because we were not 
able to reliably determine the sky values in our images, we did not attempt
to determine total magnitudes for NGC 925. 

        The photometric solutions for the line and continuum images were 
calculated accounting for the airmass of the observations 
and the response of the line and continuum filters as determined for our observations of 
the spectrophotometric standard.  Using the respective exposure times and the 
fractional contribution of the two [NII] lines to the H$\alpha$ image, we 
then can get the continuum subtracted H$\alpha$ image.  Based on the spectra 
in Martin \& Roy (1994), we assumed the H$\alpha$ line accounted for 81\% 
of the emission in the line filter, with the two [NII] lines accounting for 
14\% and 5\% of the emission.  We applied these solutions to each filter and each 
side of the galaxy separately.  The line and continuum images were aligned to 
better than 1 pixel with similar, but not exactly the same, seeing conditions.  

        Once each image was fully calibrated, we aligned and averaged together 
the east and west images of NGC 925 to form a total image of the galaxy.  To align the 
images, we identified stars present in both images, calculated the average shift 
needed to bring the two sides into alignment, and shifted the images using {\it{imshift}}.  
Images were aligned to better than 1 pixel in each dimension.  A similar procedure was 
followed to align the full galaxy B, V, R, and H$\alpha$ images to each other.  
At this point we began our analysis.  Figure 1 shows our combined B, V, R image of NGC 925.

\section{Global Photometric Properties}

One of the key thrusts of investigations of late-type spirals is the degree to which their disks are 
lopsided.  Extreme late-types such as the Magellanic spirals have dynamical centers that are significantly 
offset from the center of the outer isophotes (de Vaucouleurs \& Freeman 1972, Odewahn 1996).  We can measure 
a photometric center from our mosaiced image of NGC 925 and a dynamical center from the HI velocity field (Paper 1).  
NGC 925 is also a barred galaxy, so we can compare the photometric and dynamical center with the apparent 
center of the bar.  Magellanic spirals in particular host bars that are greatly offset from the photometric 
{\it{and}} dynamical centers.

We chose to use the isophote fitting routine in the IRAF package, {\it{ellipse}}.  We first removed the 
stars and saturated columns from the image by replacing the central regions of bright stars within a radius of 
2.4$^{\prime\prime}$ with the mean from an annulus with inner radius of 3$^{\prime\prime}$ and width 
1$^{\prime\prime}$.  While some stars were still visible in the image the residuals were only slightly 
higher than the background level of the galaxy.  We fit a series of elliptical annuli 0.$^{\prime\prime}$4 
in width to the galaxy.  The semi-major axis of each annulus was 4$^{\prime\prime}$ larger than the previous 
one.  The task ceased to produce fits which converged at a radius of 5$^{\prime}$, near the edge of our field of 
view.  We present the results of our elliptical fits in Figures 2 \& 3.  Note that the scale on the surface brightness 
plot comes from our photometric calibrations of the background sky level (although this may not be the
true sky).

A number of features stand out in the isophotal fits in Figure 2.  Overall, the results of the fits are 
very similar in all bands even though the center, position angle, and ellipticity of the galaxy all 
vary widely as a function of radius.  Within a radius of about 60$^{\prime\prime}$ (i.e. the bar region) 
the center is fit consistently in all bands.  Simply taking the means from each band we define the bar center 
to be: 2$^h$24$^m$17.0$^s \pm$0.2$^s$, 33$^o$21$^{\prime}$15$^{\prime\prime} \pm$1$^{\prime\prime}$.  The epoch for all
of the coordinates in this paper is 1950, unless otherwise stated.  
There is a sharp change in the fitted declination of the center just beyond a radius of 60$^{\prime\prime}$, 
coincident with the position of the prominent spiral arm on the southern side of the galaxy.  The
width of the ``bump'' in the fitted 
declination of the center is consistent with the width of the arm.  The fitted declination then returns 
to a value comparable with that derived from the innermost parts of the galaxy.  So if we were to ignore the 
effect of the spiral arm we find that the fitted photometric center of NGC 925 is only slowly varying with 
radius out to $\sim170^{\prime\prime}$ which is close to edge of the optical disk.  Beyond about 
200$^{\prime\prime}$ the center, position angle, and ellipticity remain relatively constant.  While the three 
bands return inconsistent fits for the center of the galaxy, we take the mean center derived from fits to the 
annulus 210$^{\prime\prime}$-270$^{\prime\prime}$ to be the center of the outer isophotes.  This yields an 
isophotal center at 2$^h$24$^m$14.8$^s \pm$0.6$^s$, 33$^o$21$^{\prime}$23$^{\prime\prime} \pm$2.5$^{\prime\prime}$ 
which is different from the bar center.  Studies of lopsided galaxies usually use 
the center of the outermost isophotes to define a photometric center of the galaxy (e.g. Odewahn 1991).  
NGC 925 shows that this center will be a function of the band with which one chooses to carry out the 
photometry.

The fitted position angle of NGC 925 in Figure 2c slowly increases from -80$^o$ to -55$^o$ from 
the center to a radius of 180$^{\prime\prime}$; beyond that it drops abruptly.  Similar trends are seen 
in Figure 2d where we plot the fitted ellipticity as a function of radius.  It is slowly decreasing from the 
center out to 180$^{\prime\prime}$ where it also drops abruptly before leveling off.  We list our adopted values 
for the center, position angle, and ellipticity in Table 3

In figure 3a we plot the azimuthally averaged surface brightness distribution for NGC 925.
We fit a double exponential to this distribution and we show the residuals in 
Figures 3b and 3c, respectively.  Residuals from our fit solely to the exponential outer-disk 
(Fig 3b) clearly illustrate the presence of a second exponential at radii within 
60$\arcsec$, and a systematic surface brightness enhancement between 
140$\arcsec$ and 180$\arcsec$.  The latter is much clearer after we subtract our 
exponential fit to the bar and re-scale the residuals in the lower panel.  We 
chose to fit an exponential to the bar, both because it represents the data 
well, and because Elmegreen \& Elmegreen (1985)  found that late-type galaxies 
including NGC 925 have exponential bars.  Both of our fits are shown in the top 
panel.  The derived scale-lengths and central brightnesses for the bar and disk are 
listed in table 3.  One very interesting property illustrated by these scale-lengths is that
the disk tends to get redder with increasing radius, as the B scale-length is much smaller than the
R scale-length.  This is contrary to what is found for most field galaxies (Vennik {\it{et al.}} 1996), and
could be indicative of a differing star formation history for NGC 925 as compared to most other galaxies.
This could also indicate the presence of a very young inner disk and bar in NGC 925.   

By looking at all the results of our isophotal fitting, we can pick out some 
distinct features in NGC 925 which are evident in both a contour plot of the 
model (figure 4), and in the graphs of the fit parameters (figures 2 \& 3).  The inner bar 
region, within 30$\arcsec$, is characterized by widely varying values for the 
center, position angle, and ellipticity that are most likely due to patchy star formation 
and dust in the center of the bar.  We get consistent results for the rest of 
the bar. The southern arm strongly affects the fits in the remainder of the 
disk.  This is evident from the drastic change in center, position angle, and 
ellipticity at 180$\arcsec$ where the spiral arm fades (as shown in the surface 
brightness fits).  The slow increase in position angle with increasing radius 
illustrates the change from the bar dominated region of the galaxy to the region 
where the spiral arm dominates.  Other variations in the fits and surface 
brightness residuals can be explained by the small scale variations in 
brightness caused by H~II regions and patchy dust within NGC 925. 

The bar of NGC 925, based on our fits shown in figure 3, extends out to approximately 60$\arcsec$ (2.7 kpc).
It is at this point where the exponential profile of the bar begins to meet the 
exponential disk.  The bar is centered at 2$^h$24$^m$17$^s$ in right ascension, 
and 33$^o$21$\arcmin$15$\arcsec$ in declination, which is not coincident with the center of 
the outer isophotes.  This is the mean from the isophotal fits of the center within the inner 60$\arcsec$. 
The bar has an ellipticity of $\sim$0.69 (b/a = 0.3), and, with a position angle of -75$^o$, it is aligned 
within a few degrees of the major axis of the galaxy.  The alignment of the bar position angle with the 
galaxy position angle, and the origination of the spiral arms from the end of the bar may indicate that the
bar has had some effect on the disk of the galaxy.  

We now have four centers derived for NGC 925: the peak of the brightness 
distribution (from the RC3), the dynamical center (from Paper I), the center of
the bar, and the center of the ``outer'' isophotes (from above).  In figure 5,
we compare the locations of these centers as a measure of the asymmetry in the
galaxy.  The size of the boxes are representative of the fitting errors; the
ellipse is the beam size from the H~I data used to model the rotation curve.  
We find that the center of the brightness distribution is nearly coincident with the 
center of the bar, while the dynamical and outer isophote centers are coincident with each other.  
The bar center is NOT coincident, however, with either the dynamical or outer isophotal center.  
This is indicative of an asymmetric galaxy and is quite similar to what is seen in SBm's such as NGC 4618
(Odewahn 1991) which has a 662 pc displacement between the bar center and the center
of the outer isophotes.  For NGC 925 the dynamical and outer isophotal centers are offset 
parallel to the bar, whereas most SBm's show bars offset perpendicular to their major axis (Odewahn 1991).  
NGC 925 may be exhibiting a different type of phenomenon, but this property of SBm's requires further study.

While we were unable to calculate total magnitudes for NGC 925, we used published values for M$_{B}$ to
examine mass-to-light ratios for NGC 925.  From the RC3, we find that NGC 925 has an M$_B$=-19.9 mag.  Using
the total and H~I masses we derived in Paper I,  we calculated M$_{tot}$/L$_B$ to
be 5.0, and M$_{H~I}$/L$_{B}$ to be 0.36, consistent with what we expect from its classification as an Scd 
spiral (Roberts \& Haynes 1994), but also similar to values found for later-type spiral galaxies.

\section{Star Formation in NGC 925}
\subsection{Global Properties}

In a seminal paper, Kennicutt (1989) observationally examined the dependence 
of star formation on the properties of the gas in the galaxy.  For his study he used the 
expression for the critical density for a instability to grow in a thin isothermal gas 
disk:
\begin{equation}
\Sigma_{crit} = \alpha \frac{\kappa c}{3.36 G}
\end{equation}
where c is the velocity dispersion of the gas, G is the gravitational constant, 
$\alpha$ is a dimensionless constant near unity which accounts for the deviation of the disk from
a thin, isothermal gas disk, and $\kappa$ is the epicyclic frequency given by:
\begin{equation}
\kappa = 1.41 \frac{V}{R}(1+\frac{V}{R}\frac{dV}{dR})^\frac{1}{2}
\end{equation}
where V is the rotation velocity at a radius R.  In our case the rotation curve is given by a Brandt
rotation curve derived from a fit to the H~I data in paper I:
\begin{equation}
V(R)=\frac{V_{max}}{R_{max}}\frac{R}{[\frac{1}{3}+\frac{2}{3}(\frac{R}{R_{max}})^n]^\frac{3}{2n}}
\end{equation}
where V$_{max}$=118 km s$^{-1}$, R$_{max}$=17.8 kpc, and n=1.46 (paper I).  
Kennicutt examined how star 
formation compared to the gas density and the expected critical density within a 
collection of galaxies.  He found that star formation, 
as traced by H$\alpha$ emission, was correlated with the gas surface density (as 
traced by H~I emission) and that star formation was suppressed when the gas surface 
density fell below a critical value.

Based on data from paper I, and using our H$\alpha$ images, we carried out a 
similar analysis on NGC 925.  From the H~I data we derive an average velocity 
dispersion, c, of 10 km s$^{-1}$ for NGC 925.  The value of $\alpha$ should be close to 
one, because galaxies should be close to being thin, isothermal disks, although Kennicutt 
(1989) finds an $\alpha$ of 0.67 fits his data better.  In figure 6 we show the critical 
surface density ($\Sigma_{crit}$) for three values of $\alpha$ and surface density of H~I 
multiplied by 1.47, $\Sigma_{gas}$, (following the example of Kennicutt 1989 to account for 
the molecular component of the gas).  We also plot the H$\alpha$ emission normalized to its 
peak.  The H$\alpha$ and H~I surface densities shownare azimuthal averages.  We find the 
H$\alpha$ surface brightness to be relatively constant throughout the galaxy before cutting 
off abruptly at about 15 kpc, whereas the gas surface densitydecreases slowly with radius 
out to 18 kpc.  This suggests that star formation in NGC 925 is not directly correlated with the 
gas surface density on a global scale.

While the abrupt decline of H$\alpha$ surface brightness occurs at the edge of our field of 
view, we can use the value of $\Sigma_{gas}$/$\Sigma_{crit}$ at this point to yield $\alpha$.  
Looking figure 6, we see that a value of $\alpha$ of near 0.3 works best for NGC 925 as opposed to 
$\alpha$=1 and $\alpha$=0.67.  In reality, the H$\alpha$ emission may continue beyond the edge 
of our field of view, implying a lower value of $\alpha$ still.  Regardless of the extent
of H$\alpha$ emission beyond the region we imaged, the star formation criterion found by Kennicutt 
(1989), $\alpha$=0.67, does not hold for NGC 925; instead, a value of $\alpha$ closer to 0.3 is more 
appropriate.  This has been found to be the case for other galaxies as well.  Values of 
$\alpha$ around 0.3 have been found for dIrr galaxiesby Hunter, Elmegreen, \& Baker (1998), and for 
low surface brightness dwarf galaxies by van Zee {\it{et al.}} (1997).  This is evidence that star 
formation in NGC 925 is governed more by local conditions than by global instabilities; a property 
more typical of later type galaxies than NGC 925.  This is not surprising as star formation can 
occur in any type galaxy even when such a criterion is not met (see Ferguson {\it{et al.}} 1998 for 
a discussion).  Ferguson {\it{et al.}} find H~II regions outside the cited limits to the optical 
disk in a sample of spiral galaxies.  Their outer H~II regions, and ours, lie on coherent spiral 
arms even at large radii.  

A better analysis of star formation criteria in NGC 925 could be done using CO data to trace the molecular 
component of the gas and using a larger field-of-view CCD to better trace the full extent of the H$\alpha$ 
emission in NGC 925.

\subsection{Star Formation in the spiral arms of NGC 925}

Comparing the H~I distribution of Paper I with the H$\alpha$ and broadband images 
of NGC 925 allows us to examine the nature of star formation in the galaxy. 
The star formation in NGC 925, as traced by the H$\alpha$ emission, is found
in two main regions.  The largest star forming regions are 
located at the end of the southern spiral arm and in the bar.  In addition, star 
formation occurs at a lower level throughout the southern spiral arm (although it may
be partially obscured by dust).  The northern arm has sporadic star formation 
near the bar, but very little elsewhere, except, perhaps, near the end of the
arm on the east side of the galaxy.  The lack of coherent star formation along
the north arms is not surprising given the weak definition of the stellar and gaseous arms. 

Figure 7 shows the H$\alpha$ in red on the H~I in blue.  In general, the star formation is occurring 
on top of H~I peaks and the H$\alpha$ emission traces the H~I spiral arms very closely.  
The coincidence between the two distributions is highly correlated across the entire galaxy, 
but the exceptions are quite interesting.  Specifically, there is a sizable 
H~II region on the northwest spiral arm near the bar which is sitting in a local H~I 
depression.  This is the only major H~II region which is not associated with a H~I peak.  
This could be due to the star formation depleting the H~I in this region, ionizing it, or 
the H~I hole could be filled with molecular gas.  

Looking at the broadband BVR image with the H~I contours (figure 8), we see
that the peak H~I does trace the optical emission (as well as the H$\alpha$ emission), 
as expected.  This figure clearly shows that the southern spiral arm is much 
stronger and more coherent than the northern arm in H~I and optical emission.  
The north side of the galaxy shows more flocculent spiral structure and less intense star
formation as compared to the south.  The southern arm also has a prominent gap in optical 
emission in the middle of the arm, while it still has a large amount of H~I.  
This may be due to a large dust lane crossing the arm at this position.  These are 
not caused by inclination, as the southern part of the galaxy is the far side of the galaxy.  
You can also see the bluest regions of the galaxy are those where large H~II regions exist, 
specifically at the end of the southern arm, and to the northeast of the bar (which is the end 
of one of the northern arms).  

As our data is of sufficient resolution and photometric quality to address the
question of triggering of star formation in spiral arms of late-type galaxies, we took cuts through 
various regions of the spiral arms and bar in NGC 925 (figure 10) to look for offsets 
between the peaks of emission in B, R, H$\alpha$, and H~I (see Beckman \& Cepa 1990).  
Figure 9 shows the location of these cuts.  These cuts were 
4$\arcsec$ wide and of varying length (as shown in figure 9).  They are 
roughly centered on the arms or bar.  The emission is an average over the width of the cut and is 
normalized to the peak intensity of that band in the cuts.  

Triggering is defined as an increase in the 
star formation efficiency in the arm versus the inter-arm region; it is not 
simply an increase in total star formation.  An offset in the peaks would suggest
there is an age gradient across the cut (in a dust free world).  Beckman \&
Cepa (1990), for instance, see such a color gradient in NGC 7479 between the B and I 
bands.  Dust, however, can produce a similar gradient due simply to an
extinction gradient within the spiral arm (Elmegreen 1995).  In addition,
we can not determine if triggering is actually occurring in the 
arm solely from an age (or color) gradient; it is entirely possible to see a 
gradient without any triggering (cf. Elmegreen 1995).  What we can say is that 
the absence of an offset between the peaks in B and R implies the lack of an 
age gradient in the underlying stellar population.  

Cuts along the southern arm (figure 10a-d) show the optical emission is well-aligned with the
H~I across the arm for most of the length of the arm.  Peaks in B and R
are well-aligned. In general, the H$\alpha$ peaks correlate with the
broadband peaks, but not always with the H~I peaks.  There are regions with no 
H$\alpha$ peaks, but with strong broadband emission.  In addition, there are two noticeable 
instances of H$\alpha$ peaks in H~I depressions.  These occur at $\pm$ 40$^{\arcsec}$
in cut c and in cut d at 40$^{\arcsec}$.  Star formation in the
south arm is located in large H~II regions which tend to lie on the peak of the H~I distribution.

Cuts across the north arms (figure 10e-g) of NGC 925 show a decidedly different phenomenon 
than those of the south arm.  Cut e shows optical emission peaking in 
an H~I depression at -50$\arcsec$.  There is also some H$\alpha$ emission
peaking with the H~I at 20$\arcsec$.  Other cuts across the north arms, such as
cut f, show a correlation between the optical and gas peaks, while cut g 
shows an offset between the optical emission and the H~I peaks.  In the north arms,
the star formation appears less prolific than in the southern arm and occurs in smaller
H~II regions which are less coherently placed than in the south of NGC 925.
This suggests that the nature of the spiral pattern, and hence the star formation, in the
north and south arms of NGC 925 are different, if not fundamentally, then in 
our perception of it.  This type of asymmetry, with one spiral arm being dominant
in a galaxy, is quite typical in later-type spiral galaxies, such as SBm's
(Odewahn 1991, de Vaucouleurs \& Freeman 1972).
NGC 925 illustrates that these properties are not limited to SBm's, but are evident
in earlier type spiral galaxies, although less prominently.  

\subsection{Star Formation in the bar of NGC 925}

In general, LTBS tend to have gas-rich bars with star formation occurring 
throughout the bar (Phillips 1995).  Friedli \& Benz (1995) suggest that 
star formation occurs along the major axis of dynamically young bars, and is concentrated in 
the center or in rings of older bars.  NGC 925's bar is quite gas-rich and has star formation 
occurring all along its major axis.  Thus, the Friedli \& Benz (1995) model would imply that NGC 
925 has a young bar.  The extremely blue center of NGC 925, compared with its outer region (see 
section 3), could indicate a dynamically young bar as well, one which has just recently started 
prolific star formation.  On the other hand, it is possible that weak bars, such as NGC 925's, are
simply inefficient at funneling gas into the galactic nucleus.  Furthermore, because corotation 
is far out in the disk (Elmegreen {\it{et al.}} 1998), the spiral arms have a large reservoir of 
gas to drive towards the center of NGC 925.  As there is no inner Lindblad resonance in NGC 925 
(Elmegreen {\it{et al.}} 1998) where the gas would pile up, the bar would be well-supplied with 
gas for long-term star formation, and it may not be young at all.   

The star formation in the bar appears to be offset somewhat from the H~I peaks. 
Examining the bar region in figure 7, we find that there is a great deal of star 
formation on the north side of the bar that is systematically displaced away from the 
H~I peaks.  This is better illustrated by cuts h-k in figure 10.  Furthermore, the minor axis cuts (i-k) 
show that the B and R peaks are systematically offset to the north of the H~I peak.  While there
is a H$\alpha$ peak aligned with the H~I peak in cut j, the main optical 
emission is offset.  This could be explained by dust extinction, depletion of the H~I by star
formation or ionization, the gas being in molecular form, or the bar being a wave phenomenon.

For this offset to be an extinction feature, some process must pile up the dust preferentially
along one side of the bar.  Dust lanes have been found in the bars of other galaxies, and their coherence
is usually attributed to shocks occurring in these bars (see Athanassoula 1992
for a nice discussion).  The bar of NGC 925, 
however, does not have strong streaming motions (paper I) which would be 
indicative of shocks. The dust appears to be in filaments in the bar region, and not 
in well defined lanes.  Furthermore, some of the cuts perpendicular to the major axis
of the bar do show star formation correlated with the H~I peaks, but all of the cuts
indicate that the stellar emission is downstream of the H~I peaks.  Extinction should affect
both H$\alpha$ emission and R band emission to a similar extent.  As this is not evident in the cuts, 
it is improbable that dust is causing the apparent offset.  

The major axis cut (h) shows the H$\alpha$ peaking throughout the bar, with the
gaps in emission likely due to the prolific dust content of the bar.  However, the 
broadband optical peaks are concentrated in a H~I depression.  This
may be due to the recent star formation and young stars evacuating or ionizing the H~I in 
this region of the bar.  The H~I depression in the bar could be filled with molecular gas, and
not truly be depleted of all gas.  Along the major axis of the bar the H~I seems to be piled up at the 
bar ends, particularly near the dynamical center of the galaxy at 
50$\arcsec$.  Spectra of the bar taken with the DensePak instrument on WIYN are
in hand and should yield insight into the nature of the ionized gas in the bar.
Higher spatial resolution H~I data will also help to untangle the dynamics of
gas in the bar.
 
If the offset of H~I south of the H$\alpha$ in the bar is not caused by dust, or by young stars ionizing or otherwise 
clearing out the H~I in the bar, or by the gas being in molecular form, then the bar could be a wave phenomenon.
This may be linked to the offset dynamical center of the galaxy as reported 
in Paper I. The dynamical center is located on the west end of the bar, so the 
geometry and dynamics make sense for this scenario, with the star formation occurring 
downstream of the gaseous peaks and the stars even further downstream,
provided that star formation is occurring as the gas moves through the wave  

\section{Discussion \& Conclusions}

We have observed NGC 925 using the WIYN telescope in B, V, R, and H$\alpha$ filters
to better understand the stellar distribution and star formation properties of the galaxy.
We have compared these observations with previously described H~I observations (Paper I)
to get a nearly complete picture of the properties of NGC 925.  

The global properties of NGC 925 are typical for a late-type spiral galaxy.  It has an obvious
bar, and two large spiral arms.  The southern spiral arm is apparently much stronger than the
northern arm, which is quite flocculent.  The relative dominance of the southern arm is apparent
both optically and in the H~I (paper I).  This difference between the northern and 
southern spiral arms is not the only asymmetry present in the galaxy.  The center of the 
galaxy, as derived from the outer isophotes, is coincident with the dynamical center (Paper I), but not with
the center of the bar.  These properties are typically associated with barred Magellanic
spirals (de Vaucouleurs \& Freeman 1972), although in the case of SBm's the offset bar and single, dominant
spiral arm tend to be much more pronounced than they are in the case of NGC 925.  Nevertheless, we see that
such properties are not limited to Magellanic spirals, but are present in more subtle ways in earlier-type
spirals.  The optical properties of NGC 925, such as mass-to-light ratio and absolute magnitude, are as would 
be expected for a typical late-type galaxy.  

From our isophote fitting, we find that NGC 925's surface brightness distribution can be characterized by a 
double exponential; one for the bar and one for the outer disk.  The bar is quite easily distinguished from 
the rest of the galaxy in the fits, by having a well-defined center, position angle, and ellipticity differing
from the outer disk.  The bar also has a exponential brightness distribution with a scale-length much
smaller than the rest of the galaxy.  There is little evidence that the bar has any effect on the structure of 
the rest of the galaxy.  The isophote fits show quite a bit of structure outside of the bar region.  This
is almost certainly due to the bright southern spiral arm, which shows up clearly as a surface brightness
enhancement in the fits.  At the same radius as this enhancement, we can see distinct changes in the fits
for the center, ellipticity, and position angle.  The fits do not seem to be affected by the weaker northern
arm.  This is further evidence for NGC 925's similarity to the SBm one-armed spiral pattern.  The fits
also show that NGC 925 gets redder with increasing radius, atypical of normal spiral galaxies in the field.  
This could indicate a recent enhancement of star formation in the inner galaxy.

From our analysis of the critical density for massive star formation in NGC 925,
using the technique of Kennicutt (1989), we find that the galaxy has widespread star formation occurring
outside of the radius where it should be suppressed.  NGC 925
has a very flat H$\alpha$ surface brightness distribution, with emission extending all the way
to the edge of our image at 15 kpc.  Kennicutt (1989) using a sample of 15 Sc galaxies found that star formation was
suppressed when $\alpha$=$\Sigma_{gas}$/$\Sigma_{crit}$ fell below 0.67.  For NGC 925 star formation continues past
this radius and is more consistent with an $\alpha\sim$0.3, like what was found by Hunter {\it{et al.}} (1998) and 
van Zee {\it{et al.}} (1997) for later-type galaxies.  The exact value of $\alpha$ is unclear, but probably lower, as the
H$\alpha$ emission could continue beyond our field of view.

As mentioned above, the spiral arms in NGC 925 are quite different.  The north arm is quite weak, 
lacks coherence, and has only patchy star formation.  The southern arm, on the other hand, can be 
traced all the way from the bar out to the edge of the galaxy and has plenty of star formation.  
Aside from a small gap, the arm is quite coherent.  There is a long dust lane associated
with the southern arm, which probably obscures some star formation.  If there is dust along 
the northern arm, it is not nearly so well-behaved.  Taking cuts across the northern and southern 
arms, we can measure the relative positions of the stars, H~II regions, and H~I.  In the southern 
arm, the strongest H$\alpha$ and stellar emission is closely aligned with the peak of the H~I.  
Large scale coherence breaks down in the northern half of the disk.  At some places 
(40-50$^{\prime\prime}$ in cut f) star formation is well correlated with peaks in the H~I distribution.
At other locations ($\sim$40$^{\prime\prime}$ in cut g) star formation is
displaced off of the H~I peaks while elsewhere the H~I and H$\alpha$ are
anti-correlated (e.g. $\sim$50$^{\prime\prime}$ in cut e).  
The difference between the north and south arms suggests that there may be some fundamental difference 
in the nature of the two arms.  

Using the same cuts, we looked for evidence of triggering by the spiral arms.  If there is triggering 
in the arms, we might expect a color gradient (implying an underlying age gradient) perpendicular to 
the arms.  Such a color gradient can also be caused by dust extinction.  The lack of a color gradient 
would imply the lack of an age gradient, and, hence a lack of triggering.  While the cuts of the 
north arms show that the stellar emission is typically offset from the
H$\alpha$ and H~I emission, there is no offset present between the B \& R peaks.  Similarly, the 
southern arm shows no offset between the broadband emission peaks.  The lack of an offset shows that 
there is no color gradient, and hence it is unlikely that an age gradient due to triggering in the 
spiral arm is occurring.  This suggests that there is no larger scale organization of the star formation 
in the northern disk--a result reminiscent of irregular galaxies. Star formation
must be driven by local conditions and processes in the northern half, while
larger scale phenomena organize star formation in the bar and southern arm.  

Finally, we examined the nature of the NGC 925's bar and the star formation occurring within it.  As is expected for 
late-type bars (Phillips 1995), NGC 925 has star formation occurring all along the bar.  Cuts perpendicular to the 
major axis of the bar show that the star formation is occurring along the major axis, but the broadband optical
emission is systematically offset to the north of the H~I peaks.  While this could be a dust effect, we would expect
the H$\alpha$ emission to be similarly affected, but it is not.  This could also be due to the massive star formation 
in the bar ionizing or clearing out the H~I in the bar.  The systematic offset may be related to the dynamical center 
being offset from the bar center, and hence the bar could be a wave pattern moving through the gaseous medium in NGC 925. 
We might expect to see the peak of the stellar emission downstream of the star formation, as is the case here.
It may be that the bar of NGC 925 is not actually a bar, but has more in common with the spiral arms of NGC 925.  

Overall, this study has shown that NGC 925 is a prototypical late-type spiral galaxy, with some properties that are 
characteristic of Magellanic spirals, such as an off-center bar and a dominant spiral arm.  These traits are less 
pronounced in NGC 925 than in an SBm indicating a probable smooth transition of these properties
from Scd galaxies to Sm's.  Massive star formation in NGC 925 persists beyond the radius predicted by Kennicutt (1989) 
for late-type spirals, but behaves more like dIrr's (Hunter {\it{et al.}} 1998).  
This result may be exacerbated by wider field imaging of NGC 925, however, and star formation almost certainly 
continues at a low level beyond this point (see Ferguson {\it{et al.}} 1998 for an example).  The north and south arms 
are not only different in brightness, but in coherence and in the distribution of stars across the arms.  The north arm 
has stars offset from H~I peaks, while the south arm has both star formation and stellar emission coincident with the
neutral gas peaks.  Finally, the bar of NGC 925 appears to be a typical late-type bar being gas-rich and having 
star formation throughout, but the stars are offset from the H$\alpha$ and H~I peaks.  This fact coupled with the
offset dynamical center of NGC 925 suggests that the bar may be more of a wave phenomenon, similar to a spiral arm.  

\acknowledgements
The authors would like to thank the staff at the WIYN observatory and on Kitt Peak for helping make our 
observing run go smoothly, and for their excellent assistance in producing the wonderful data from the
WIYN telescope.  The authors also thank the anonymous referee for comments improving the quality of this paper.
This work was partially supported by NSF grant AST 96-16907 to E.M.W.  
This research has made use of the NASA/IPAC Extragalactic Database (NED) which is operated by the Jet 
Propulsion Laboratory, California Institute of Technology, under contract with the National Aeronautics 
and Space Administration.  

\vfil\eject

\vfil\eject

\centerline{\bf{Figure Captions}}

\figcaption{} A combined B,V,R image from the WIYN 3.5m telescope.  B is represented by 
blue, V by green, and R by red.  

\figcaption{} The results from the isophote fitting of NGC 925.
Counter-clockwise from upper left they are Right Ascension of center, Declination of center, ellipticity, 
and position angle.  Blue represents fits to the B band image, Green for V, and Red
for R.

\figcaption{} The results of the isophote fits to the surface brightness in
B,V,R for NGC 925.  The colors are the same as figure 2. The upper panel shows
the data on an arbitrary scale (points) and the fits to the bar and disk
(solid lines).  The middle panel shows the residuals from just the disk fit.  
The lower panel shows the residuals from both the bar \& disk fits.  The solid
line in the lower 2 panels mark the zero level.  

\figcaption{} A representation of the model isophotes as fit to the R band image of NGC 925.  
Ellipses are plotted for one out of every three annuli.  They illustrate the relative location
and orientation of each annulus, but do not reflect the surface brightness distribution.  

\figcaption{} This plot shows 4 centers of NGC 925: the center of NGC 925 according to the RC3 (RC3), 
the center of the bar (B), the center of the outer isophotes (ISO), and the dynamical center (DYN).  The boxes
represent the derived errors to the fitted centers.  The ellipse is the FWHM
of the beam from the H~I observations in paper I. 

\figcaption{} a) Top panel is a plot of gas surface density (solid line), as measured by H~I
observations from paper I, and multiplied by 1.47 to account for molecular
gas. The dashed line is the critical gas surface density for collapse an $\alpha$=1.  The
dot-dash line is the radial H$\alpha$ distribution normalized to its peak.
The vertical dotted line indicates the nearest edge of the chip from the center of
the bar of NGC 925.  The bottom panel indicates the ratio of the measured gas surface
density to the critical density.  The horizontal dotted line is the level when
massive star formation should cease.  The vertical horizontal line is the same
as in the top panel.  b) same as in a, but for $\alpha$=0.67.  c) same as a, but for 
$\alpha$=0.3.

\figcaption{}  This figure shows the H$\alpha$ emission from NGC 925 (in red)
and the H~I emission (in blue) overlaid on each other.  Regions that are white
are where there are peaks in both H$\alpha$ and H~I.  

\figcaption{}  This figure shows the H~I contours from paper I (in yellow)
overlaid on the combined B,V,R WIYN image from figure 1.  Hatched contours
indicate a H~I depression.  The H~I beam is shown in the lower right.

\figcaption{}  This is an optical R band image of NGC 925, with the location
of the cuts marked on it.  The rectangular boxes are show the region over
which the cut was taken.  The letters correspond to the cuts shown in figure 10, and
also marks the most positive point in the cut. 

\figcaption{}  The results of the cuts across the southern arm, northern arms, and bar of
NGC 925.  Each band of observations is normalized to its peak.  The solid line is B band, the dashed
line is R, the dot-dash line is H$\alpha$, and the dotted line is H~I column density.  Cuts a-d are
for the southern arm.  Cuts e-g are for the northern arms.  Cut h is the bar major axis, and cuts i-k are 
parallel to the bar minor axis.  The most positive points in the cuts are marked by the letters in figure 9.

\end{document}